\documentclass[9pt,twocolumn,twoside]{pnas-new}

\templatetype{pnasresearcharticle} 

\setboolean{displaywatermark}{false}
\usepackage{physics}
\usepackage{amsmath}
\usepackage{caption}
\usepackage{subcaption}

\title{Band gaps of crystalline solids from Wannier-localization based optimal tuning of a screened range-separated hybrid functional}

\author[a]{Dahvyd Wing}
\author[a]{Guy Ohad}
\author[b,c]{Jonah B. Haber}
\author[d]{Marina R. Filip}
\author[b,c]{Stephen E. Gant}
\author[b,c,e]{Jeffrey B. Neaton}
\author[a,1]{Leeor Kronik}

\affil[a]{Department of Molecular Chemistry and Materials Science, Weizmann Institute of Science, Rehovoth 76100, Israel}
\affil[b]{Department of Physics, University of California, Berkeley, Berkeley, California 94720, USA }%
\affil[c]{Materials Sciences Division, Lawrence Berkeley National Laboratory, Berkeley, California 94720, USA}
\affil[d]{ Department of Physics, University of Oxford, Oxford OX1 3PJ, United Kingdom }
\affil[e]{Kavli Energy NanoSciences Institute at Berkeley, Berkeley, CA, 94720, USA }

\leadauthor{Dahvyd Wing}

\significancestatement{Density functional theory (DFT) is successful in predicting a wide variety of material properties, but its use for electronic and optical properties has been hampered by its poor prediction of the band gap of a material. Here we provide a method, within the rigorous framework of generalized Kohn-Sham theory, that can accurately predict the band gap of a material, to within experimental uncertainty, using a specific class of density functionals called range-separated hybrid functionals. This method employs an orbital localization procedure to non-empirically select system-specific parameters of the hybrid functional. We expect the approach to be useful in predicting other electronic and optical properties of interest.}

\authorcontributions{D. W., J. B. N., and L. K. designed research. G. O. and D. W. performed research, with J. B. H., M. R. F., S. E. G., J. B. N., and L. K. assisting throughout in implementation and analysis; D. W., G. O., and L. K. drafted the paper, incorporating edits from all authors.}
\authordeclaration{The authors declare no conflict of interests.}

\correspondingauthor{\textsuperscript{1} E-mail: leeor.kronik@weizmann.ac.il}


\begin{abstract}
Accurate prediction of fundamental band gaps of crystalline solid state systems entirely within density functional theory is a long standing challenge. Here, we present a simple and inexpensive method that achieves this by means of non-empirical optimal tuning of the parameters of a screened range-separated hybrid functional. The tuning involves the enforcement of an \textit{ansatz} that generalizes the ionization potential theorem to the removal of an electron from an occupied state described by a localized Wannier function in a modestly sized supercell calculation. The method is benchmarked against experiment for a set of systems ranging from narrow band gap semiconductors to large band gap insulators, spanning a range of fundamental band gaps from 0.2 to 14.2 eV and is found to yield quantitative accuracy across the board, with a mean absolute error of $\sim$0.1 eV and a maximal error of $\sim$0.2 eV.
\end{abstract}

\dates{This manuscript was compiled on \today}
\doi{\url{www.pnas.org/cgi/doi/10.1073/pnas.XXXXXXXXXX}}

\begin{document}

\maketitle
\thispagestyle{firststyle}
\ifthenelse{\boolean{shortarticle}}{\ifthenelse{\boolean{singlecolumn}}{\abscontentformatted}{\abscontent}}{}

\dropcap{T}he fundamental band gap of a semiconductor or insulator, defined as the difference between the ionization potential and the electron affinity of the material, is an essential material property. However, predicting it from first principles using density functional theory (DFT) has proven to be challenging \cite{kuemmel_kronik_2008,onida_rubio_2002}. The Kohn-Sham (KS) lowest unoccupied - highest occupied eigenvalue gap cannot be equated with the fundamental gap even if the exact (and generally unknown) exchange-correlation functional is used \cite{perdew_levy_1983,sham_schlueter_1983}. This is because the KS potential features a discontinuity (known as the derivative discontinuity) as the number of electrons crosses integer values \cite{perdew_balduz_1982}, which results in a different reference potential for electron removal and addition, and therefore typically causes KS eigenvalue differences to underestimate the fundamental band gap by as much as 50\% \cite{godby_sham_1986, allen_tozer_2002,chan_1999}. In some cases, the discrepancy can be larger and even lead to the spurious prediction of a metallic state \cite{massidda_serra_1990,Lima_2002}. For finite systems, despite this eigenvalue discrepancy one can still calculate accurate fundamental gaps from total energy differences between the cation, neutral, and anion systems \cite{tozer_de_proft_2005}. For solid state systems, the subject of this work, this total energy differences approach would work for the exact exchange-correlation functional. However, it fails for functionals without a derivative discontinuity, for which delocalization of the KS orbitals causes the total energy difference to converge with increasing system size to the KS eigenvalue difference rather than to the true fundamental gap \cite{perdew_1985,godby_white_1998,mori-sanchez_yang_2008,kraisler_kronik_2014,vlcek_eisenberg_steinle-neumann_baer_2015,gorling_2015, Perdew_gorling_2017}.

Many DFT-based strategies for obtaining the fundamental band gaps of solids by going beyond the KS scheme have been proposed over the years, e.g., Refs.\ \cite{bylander_kleinman_1990,geller_wimmer_2001, heyd_martin_2005,cococcioni_de_gironcoli_2005,anisimov_kozhevnikov_2005,ferreira_teles_2008, shimazaki_asai_2008,zhao_truhlar_2009, tran_blaha_2009, chan_ceder_2010,marques_botti_2011, Skone_Govoni_Galli_2014,gorling_2015, skone_galli_2016, ma_wang_2016, weng_wang_2017, verma_truhlar_2017, Perdew_gorling_2017, nguyen_marzari_2018,cui_jiang_2018,chen_pasquarello_2018, miceli_pasquarello_2018, bischoff_pasquarello_2019, bischoff_pasquarello_2019_perovskites, weng_wang_2020, cipriano_pacchioni_2020, tancogne_dejean_rubio_2020, lee_son_2020,lorke_frauenheim_2020}. However, two outstanding issues remain. One is achieving a level of accuracy that is on par with that of experiment (approximately 0.1 eV) for a wide range of materials, from narrow band gap semiconductors to wide band gap insulators. For example, the Heyd-Scuseria-Ernzerhof (HSE) functional \cite{Heyd_Ernzerhof_2003,*heyd_ernzerhof_2006}, a short-range hybrid, is one of the best functionals at predicting band gaps for semiconductors, but it underestimates the band gaps of insulators by as much as several eV, e.g. predicting a band gap of 6.4 eV for MgO and 11.4 eV for LiF \cite{borlido_botti_2019}. The other remaining issue is to predict the band gap non-empirically within a formally exact framework. Experience shows that if this is achieved, significant improvement in the prediction of other material properties, from defect energetics \cite{freysoldt_van_de_walle_2014} to optical absorption \cite{maitra_2016,Byun_Ullrich_2020}, is also achieved. Here, we show that both these issues can be resolved simultaneously, by overcoming the derivative discontinuity limitation within the rigorous framework of generalized Kohn-Sham (GKS) theory. This is achieved by non-empirical, system specific, optimal parameter tuning of a screened range-separated hybrid (SRSH) functional, based on enforcement of an \textit{ansatz} generalizing the ionization potential (IP) theorem to the removal of an electron from an orbital corresponding to a Wannier function.

\section*{Theory}

Our starting point is the screened range-separated hybrid (SRSH) functional \cite{refaely-abramson_kronik_2013}, which 
mixes a fraction of exact exchange and semi-local exchange, as in a standard hybrid functional \cite{becke_1993,perdew_burke_hybrid_1996}, but with a generally different fraction used in the short-range and in the long-range. This is accomplished by partitioning the exchange part of the Coulomb interaction using the identity\footnote{This is equivalent to the more compact representation in previous papers \cite{refaely-abramson_kronik_2015}, where the xx and KSx terms are combined and $\alpha + \beta = 1/\epsilon_\infty$ }
\begin{equation}
\label{eq:rsh}
\begin{split}
 \frac{1}{r} =\underbrace{ \alpha\frac{ \textrm{erfc}(\gamma r)}{r} }_{\textrm{xx, SR}}+& \underbrace{(1-\alpha )\frac{\textrm{erfc}(\gamma r)}{r}}_{\textrm{KSx, SR}}+\\ &\underbrace{\frac{1}{\epsilon_\infty}\frac{ \erf(\gamma r)}{ r}}_{\textrm{xx, LR}}  +\underbrace{\left(1-\frac{1}{\epsilon_\infty}\right)\frac{ \erf(\gamma r)}{ r}}_{\textrm{KSx, LR}}\,,
 \end{split}
\end{equation}
where the first and third terms are treated with exact exchange (xx) and the second and fourth terms are replaced with (semi)local KS exchange (KSx). $\alpha$ is then the fraction of short-range (SR) exact exchange, $\frac{1}{\epsilon_\infty}$ the fraction of long-range (LR) exact exchange, and $\gamma$ the range-separation parameter that determines the transition from short to long range with increasing $r$, where $r$ is the interelectron coordinate. The SRSH exchange-correlation functional is then given by

\begin{equation}
\begin{split}
E_{\textrm{xc}}^{\textrm{SRSH}}(\alpha, \gamma, \epsilon_\infty)=& 
\alpha E_{\textrm{xx}}^{\textrm{SR},\gamma} + (1-\alpha) E_{\textrm{KSx}}^{\textrm{SR},\gamma} +\\&\frac{1}{\epsilon_\infty} E_{\textrm{xx}}^{\textrm{LR},\gamma} \!+\!\left(1-\frac{1}{\epsilon_\infty}\right)\! E_{\textrm{KSx}}^{\textrm{LR},\gamma}\! +\! E_{\textrm{KSc}}\,,
\end{split}
\end{equation}
where KSc denotes (semi)local KS correlation. 

The SRSH functional has several advantageous features. First, being a (range-separated) hybrid functional, it is a special case of the rigorous theoretical framework provided by generalized Kohn-Sham (GKS) theory \cite{seidl_levy_1996,gorling_levy_1997,kronik_stein_refaely-abramson_baer_2012,baer_kronik_2018}. Therefore, self-consistent exchange-correlation potentials and kernels can be obtained in a straightforward manner by taking appropriate derivatives. Second, by setting $\epsilon_\infty$ to the orientationally-averaged high-frequency (ion-clamped) dielectric constant, the functional possesses the correct average long-range dielectric screening \cite{shimazaki_asai_2008, refaely-abramson_kronik_2013}. This is known to be an important criterion for producing accurate band gaps \cite{wing_kronik_2020,shimazaki_asai_2008,kronik_stein_refaely-abramson_baer_2012,refaely-abramson_kronik_2013, Skone_Govoni_Galli_2014,skone_galli_2016,chen_pasquarello_2018,manna_refaely-abramson_kronik_2018, bischoff_pasquarello_2019_perovskites,  kronik_kuemmel_2018, lorke_frauenheim_2020, tal_pasquarello_2020, cui_jiang_2018}, optical absorption spectra \cite{refaely-abramson_kronik_2015,manna_refaely-abramson_kronik_2018,wing_kronik_2019, sun_ullrich_2020_prr, sun_ullrich_2020, wing_kronik_2020_insb}, and defect energy levels \cite{wing_kronik_2020, gerosa_pacchioni_2015}. Third, the short-range exact exchange fraction, $\alpha$, can be chosen to balance exchange and correlation effects and mitigate self-interaction errors in the short range \cite{Heyd_Ernzerhof_2003,*heyd_ernzerhof_2006, luftner_puschnig_2014,egger_kronik_2014}. Here, we use semilocal exchange and correlation components based on the Perdew-Burke-Ernzerhof (PBE) functional and adopt the default value of $\alpha=0.25$, as in the hybrid (PBE0) \cite{perdew_burke_hybrid_1996, adamo_barone_1999} and short-range hybrid (Heyd-Scuseria-Ernzerhof, HSE) \cite{Heyd_Ernzerhof_2003,*heyd_ernzerhof_2006} functionals.

With values for $\alpha$ and $\epsilon_\infty$ defined uniquely, the only parameter left undetermined is the range-separation parameter, $\gamma$. For finite systems, $\gamma$ can be \textit{optimally tuned}, i.e., determined from first principles based on the satisfaction of physical criteria \cite{stein_baer_2010,kronik_stein_refaely-abramson_baer_2012}.  This has typically been achieved by selecting $\gamma$ to satisfy the IP theorem, sometimes also known as the DFT version of Koopmans' theorem or as the generalized Koopmans' theorem \cite{perdew_balduz_1982,Almbladh_von_Barth_1985,perdew_levy_1997,levy_sahni_1984}. It states that for the exact (G)KS functional
\begin{equation}
    \label{eq:IP}
    \epsilon_{ho} = E(N) - E(N-1) \equiv -I,
\end{equation}
where $\epsilon_{ho}$ is the highest occupied eigenvalue, $E(N)$ is the ground-state total energy of the system with $N$ electrons, $E(N-1)$ is the ground-state total energy of the system with one electron removed, and $I$ is the ionization potential. For finite systems, deviation from the IP theorem is equivalent to a missing derivative discontinuity, $\Delta_{xc}$, in the approximate exchange-correlation functional (as compared to the exact functional)\cite{stein_baer_2012}, such that $\epsilon_{ho} +I \approx \frac{\Delta_{xc}}{2}$ \cite{allen_tozer_2002}. Within the GKS framework, the presence of a Fock-like operator ``absorbs'' some of the derivative discontinuity, such that each parameterization of the Fock-like operator in a hybrid functional implies a different derivative discontinuity in the exact remainder multiplicative functional \cite{seidl_levy_1996,garrick_kronik_2020}. Choosing $\gamma$ to satisfy the IP theorem is therefore tantamount to selecting a functional form with a negligible missing derivative discontinuity, i.e., there is no derivative discontinuity in the exact functional that the approximate functional needs to capture. This means that the exact GKS eigenvalue gap will equal the fundamental gap for that system \cite{stein_baer_2010,kronik_stein_refaely-abramson_baer_2012, Perdew_gorling_2017}. Furthermore, a generalized Kohn-Sham scheme that yields an eigenvalue gap equal to the true fundamental gap will also yield the correct fundamental gap from total energy differences \cite{Perdew_gorling_2017}.  

While the above method, known as the optimally-tuned range-separated hybrid (OT-RSH) functional approach, has been very successful for determining the fundamental gap in molecules (see, e.g., Refs.\ \cite{stein_baer_2010,refaely_kronik_2011,kronik_stein_refaely-abramson_baer_2012, autschbach_srebro_2014,phillips_dunietz_2014, foster_allendorf_2014,korzdorfer_bredas_2014}), it is not generally helpful for determining the band gap in solids. Due to the delocalized nature of the orbital corresponding to $\epsilon_{ho}$ in solid state systems, the IP theorem is trivially satisfied for all parameterizations of the Fock operator, regardless of whether the corresponding exact remainder functional has a derivative discontinuity or not \cite{mori-sanchez_yang_2008,kraisler_kronik_2014,vlcek_eisenberg_steinle-neumann_baer_2015,gorling_2015}. Therefore, optimal tuning can not be applied without further modification\footnote{except in the special case of a molecular solid, where it could be ``inherited'' from the underlying molecule \cite{refaely-abramson_kronik_2013,manna_refaely-abramson_kronik_2018}} and predictive power is lost.

The \textit{a priori} selection of $\gamma$ in the solid state has remained an open question and an active area of research \cite{gerber_kresse_2007,miceli_pasquarello_2018,skone_galli_2016,chen_pasquarello_2018,bischoff_pasquarello_2019_perovskites}. Because it is the delocalization of orbitals that prevents the use of optimal tuning in the solid state, the next logical step is to remedy the situation by creating localized orbitals. Miceli et al. \cite{miceli_pasquarello_2018} have achieved this by introducing a point defect in a supercell of a bulk solid and enforcing the IP theorem for the orbital localized around the defect. While this approach is well-justified physically and indeed significantly improves the predicted band gaps, it still suffers from sensitivity to the type of defect used for tuning \cite{miceli_pasquarello_2018,bischoff_pasquarello_2019}. Furthermore, ideally we would like to predict the properties of the pristine crystalline material without changing it. 

An alternate route for solving the problem of optimal tuning in the solid state is to rely on a different scheme to create orbital localization. Indeed, several recent strategies for band gap estimation have relied on different kinds of localized orbitals for obtaining correction terms that compensate for the missing derivative discontinuity \cite{anisimov_kozhevnikov_2005,ma_wang_2016,weng_wang_2017,weng_wang_2020, miceli_pasquarello_2018, bischoff_pasquarello_2019, bischoff_pasquarello_2019_perovskites, nguyen_marzari_2018, li_yang_2017}. Here, we exploit Wannier functions, which are a localized orthonormal basis set obtained via a unitary transformation of a set of Bloch wavefunctions \cite{marzari_vanderbilt_2012}.  While many different sets of Wannier functions can be produced from a set of Bloch functions, maximally localized Wannier functions \cite{marzari_vanderbilt_2012} often match intuition for chemical bonds in solids and are ``natural'' localized orbitals of the bulk system. However, Wannier functions are not eigenfunctions of the (G)KS Hamiltonian \cite{marzari_vanderbilt_2012}; moreover, a system with $N-1$ electrons, the density of which corresponds to the density of the $N$-electron ground state with the charge density of a Wannier function removed, is \textit{not} the ground state of the $N{-}1$-electron system. Thus, the IP theorem does not strictly apply to the removal of an electron from a state corresponding to a Wannier function. Ma and Wang \cite{ma_wang_2016} proposed as an \textit{ansatz} that piecewise linearity of the total energy (which is equivalent to the IP theorem \cite{stein_baer_2012}) be satisfied in this case too, and this \textit{ansatz} has been used to generate correction terms for semilocal functionals which improved the accuracy of the computed band gaps \cite{ma_wang_2016,weng_wang_2017,weng_wang_2020}. Here, we adopt this \textit{ansatz}, but instead of using it to generate a correction term, we use it to select the range-separation parameter, $\gamma$, in the SRSH functional of Eq.\ (\ref{eq:rsh}). To do so, we seek a value of $\gamma$ that satisfies $\Delta I^\gamma = 0$, where
\begin{equation}
\label{eq:deltaI}
  \Delta I^\gamma=  E_{\textrm{constr}}^\gamma[\phi](N-1)- E^\gamma(N)+ \bra{\phi} \hat{H}_{\textrm{SRSH}}^\gamma \ket{\phi}.
\end{equation}
Here $E_{\textrm{constr}}^\gamma[\phi](N-1)$ is the total energy of a system with $N-1$ electrons, including an image charge correction, under the constraint that the state corresponding to a Wannier function, $\phi$, is not occupied;  $\hat{H}_{\textrm{SRSH}}^\gamma$ is the Hamiltonian of the SRSH functional with $N$-electrons. $\bra{\phi} \hat{H}_{\textrm{SRSH}} \ket{\phi}$ is the energy of the Wannier function, which we calculate via
\begin{equation}
\label{eq:wannier_energy}
   \bra{\phi} \hat{H}_{\textrm{SRSH}}\ket{\phi} = \sum_{i=1}^{\infty} \abs{\bra{\phi}\ket{\psi_i}}^2\epsilon_{i}, 
\end{equation}
where $\psi_i$ and $\epsilon_i$ are the GKS eigenfunctions and eigenvalues of the $N$-electron system, and the sum is over all orbitals. 

To calculate the total energy of the $N-1$ electron system prior to image charge corrections, $\tilde{E}_{\textrm{constr}}[\phi](N-1)$, we use constrained DFT, inspired by approaches used in DFT +U \cite{dederichs_akai_1984,nakamura_tsuneyuki_2006,hybertsen_christensen_1989}. We add a Lagrange multiplier to minimize the ground state energy under the constraint that the ``occupation'' of the Wannier function is equal to $f_{\phi}$, that is
\begin{equation}
    \label{eq:contraint}
    \begin{split}
    \tilde{E}_{\textrm{constr}}[\phi](N-1)=\min_{\psi}\ &E[\{\psi\}](N-1)\\
    &+ \lambda \left( \sum_{i=1}^{\textrm{N-1}} \abs{\bra{\psi_i}\ket{\phi}}^2 - f_{\phi} \right),
    \end{split}
\end{equation}
where $\{\psi\}$ and $E[\{\psi\}](N-1)$ are the eigenfunctions and total energy of the system with $N\!-\!1$ electrons, and $\lambda$ is a Lagrange multiplier. Taking the functional derivative of Eq.\ (\ref{eq:contraint}) then yields the constrained GKS equation:
\begin{equation}
    \label{eq:GKS}
    \hat{H}_{\textrm{SRSH}}\ket{\psi_i}+ \lambda \ket{\phi} \bra{\phi}\ket{\psi_i}  = \epsilon_i \ket{\psi_i},
\end{equation}
which we solve self-consistently for the $N-1$ electron system. $\lambda$ determines the occupation of the Wannier function, $f_\phi$, and, in this case we enforce $f_{\phi} < 4 \times 10^{-4}$ by setting $\lambda  = 15 \textrm{ Ry}$. We find that, in general, by setting $\lambda > \sim 5 \textrm{ Ry}$ the Wannier function is nearly completely unoccupied and the desired $N\!-\!1$-electron system is obtained. Due to using periodic boundary conditions, we correct $\tilde{E}_{\textrm{constr}}[\phi](N-1)$ using the Makov-Payne monopole image charge correction for a charged system \cite{makov_payne_1995,leslie_gillan_1985,komsa_pasquarello_2012}:
\begin{equation}
    \label{eq:ICC}
    E_{\textrm{constr}}[\phi](N-1)= \tilde{E}_{\textrm{constr}}[\phi](N-1) + \frac{\alpha_{mad}q^2}{2 \epsilon_\infty L},
\end{equation}
where $\alpha_{mad}=2.837$ is the Madelung constant for a simple cubic cell; $q$ is the charge of the system, e.g. $q=1$; and $L$ is the length of the supercell.

Use of the above ingredients allows us to investigate a wide range of crystalline solids, from narrow band gap semiconductors to wide band gap insulators, using a four-step procedure which we first overview and then describe in detail. In step 1, we calculate the orientationally-averaged high-frequency (ion-clamped) dielectric constant, $\epsilon_\infty$, using a primitive unit cell and use this to set the fraction of long-range exact exchange in the SRSH functional \cite{refaely-abramson_kronik_2015,skone_galli_2016, wing_kronik_2019}. In step 2, we calculate the Wannier functions of a supercell of an $N$-electron system. In step 3, we calculate $\Delta I$ using the SRSH functional with a particular value of $\gamma$. This entails calculating $E(N)$ and $E_{\textrm{constr}}[\phi](N-1)$ using the Wannier function and the supercell from step 2. We iterate step 3 for different range-separation parameters until we find the range-separation parameter that yields $\abs{\Delta I^{\gamma}} < 0.02 \textrm{ eV}$. Here and throughout we refer to the functional obtained from this choice of parameters as the Wannier-localized, optimally-tuned SRSH (WOT-SRSH) functional. In step 4, we calculate the fundamental band gap for the material with this functional, using a primitive unit cell. Importantly, no empirical fitting is introduced in any step.

For step 1, we use HSE to calculate the dielectric constant for semiconductors \cite{paier_kresse_2008}, including spin-orbit coupling effects where necessary, and PBE0 to calculate the dieletric constant for insulators (i.e. GaN and materials with larger band gaps) \cite{ Skone_Govoni_Galli_2014}. For more details, see Section I of the supplementary information (SI) \cite{SM}.

For both steps 2 and 3, we use supercells evaluated with a $\Gamma$-point-only $k$-grid, rather than a unit cell with a larger $k$-grid. Recall that in the absence of constraints, generally a unit cell calculation with an $n{\times}n{\times}n$ $k$-grid is equivalent to the calculation of a supercell containing $n{\times}n{\times}n$ replicas of the unit cell. However, the constrained removal of an electron from a Wannier function is not the same in the two cases. In the supercell, it would involve removing an electron from one well-localized Wannier function, whereas in the $k$-point sampled unit cell, it would be equivalent to removal of a fraction of an electron from each replicated Wannier function within the supercell. The latter scenario is not desirable as it would reduce charge localization. As for the size of the supercells used in steps 2 and 3, materials with cubic symmetry use a $2{\times}2{\times}2$ conventional supercell and Wurtzite materials use a $3{\times}3{\times} 2$ conventional supercell. This converges $\Delta I$ calculations to $\sim 0.02$ eV and $\sim 0.07$ eV respectively (see Section II of the SI for convergence details \cite{SM}). 

With regard to the Wannier function used in steps 2 and 3, we find that $\Delta I$ is sensitive to the character of the Wannier function used and that it is important to choose the Wannier function to be composed of the top-most valence bands, as opposed to, e.g., deeper valence states or semi-core states. To achieve that in practice, we compute the energy of each Wannier function according to Eq.\ \ref{eq:wannier_energy} and select the one with the highest energy. For each type of Wannier function, the Wannierization process produces a set of translationally/rotationally symmetric Wannier functions, one at every equivalent atomic site, all of which produce virtually identical $\Delta I$ results. We also find that $\Delta I$ is insensitive to the functional used to generate the Wannier function. We therefore use the PBE functional to generate the Wannier function, except for narrow gap semiconductors where PBE produces a spurious metallic state, in which case we use PBE0 instead.

For step 3, when calculating the image charge correction via Eq.\ (\ref{eq:ICC}), we use the orientationally-averaged $\epsilon_\infty$ calculated in step 1. For Wurtzite materials, we use $L=\sqrt[3]{\Omega}$, where $\Omega$ is the volume of the supercell. For these weakly anisotropic systems, this produces an image charge correction that is nearly equivalent to more exact methods of incorporating anisotropy \cite{rurali_cartoixa_2009}.

For step 3, we further note that in the original OT-RSH one often seeks to fulfill the IP theorem for both the $N$-electron system and the $N{+}1$ electron system, which corresponds to setting the highest-occupied molecular orbital (HOMO) and lowest-unoccupied molecular orbital (LUMO) to the ionization potential and electron affinity, respectively.
 \cite{stein_baer_2010,kronik_stein_refaely-abramson_baer_2012}. While this is useful for atoms and small molecules, experience with larger molecules already shows that (unless totally different moieties are involved) either tuning produces similar values of $\gamma$. This can be explained by the fact that the overall character of the system changes little upon electron addition or removal for larger systems. For solids, the difference between tuning for the ionization energy or the electron affinity is expected to be even smaller. It is then much more convenient to address the ionization potential alone, owing to the technical challenge of tuning using electron affinities, in that the added electron localized in the Wannier function (composed of conduction band states) may hybridize with states in the valence band. 

For step 3, when $\alpha$ is close to $\frac{1}{\epsilon_\infty}$, varying $\gamma$ does not greatly change the amount of exact exchange in the functional and the search for an optimal $\gamma$ may fail. In this case, the value of $\alpha$ is slightly increased beyond the default value of 0.25 and the process of looking for the optimal $\gamma$ is repeated (for detailed information on the {\it a priori} selection of $\alpha$, as well as the effect of different $\alpha$ values on the predicted band gap, see Section III of the SI \cite{SM}). Our $\Delta I$ calculations do not include spin-orbit coupling effects. However, in step 4 we include spin-orbit coupling effects where necessary. When calculating indirect band gaps in step 4, we either use Wannier interpolation or sample the $k$-point of the conduction band minimum.

 Finally, we emphasize strongly that enforcement of the Wannier-localizad IP \textit{ansatz} is only used for selection of functional parameters (step 3). Once this step is complete, a standard range-separated hybrid functional is obtained and the predicted band gap is read directly off its eigenvalues (step 4), without any localization. This functional can then be used, as is, for the calculation of any further material property of interest.

\section*{Results and Discussion}

\begin{figure*}[htbp!]
\begin{subfigure}{0.46\textwidth}
\includegraphics[scale=0.295]{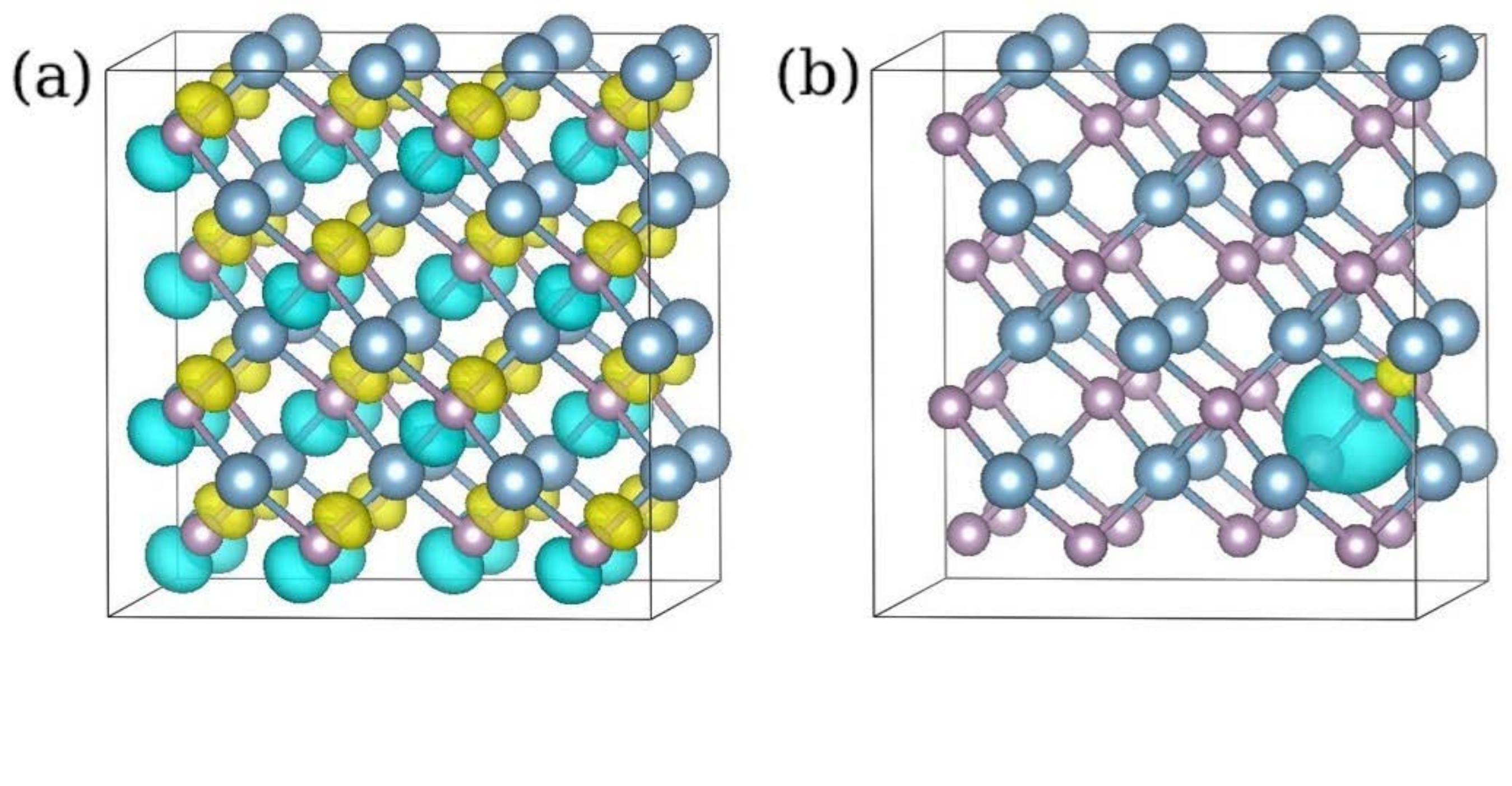}
\end{subfigure}
\begin{subfigure}{0.5\textwidth}
\includegraphics[scale=0.53]{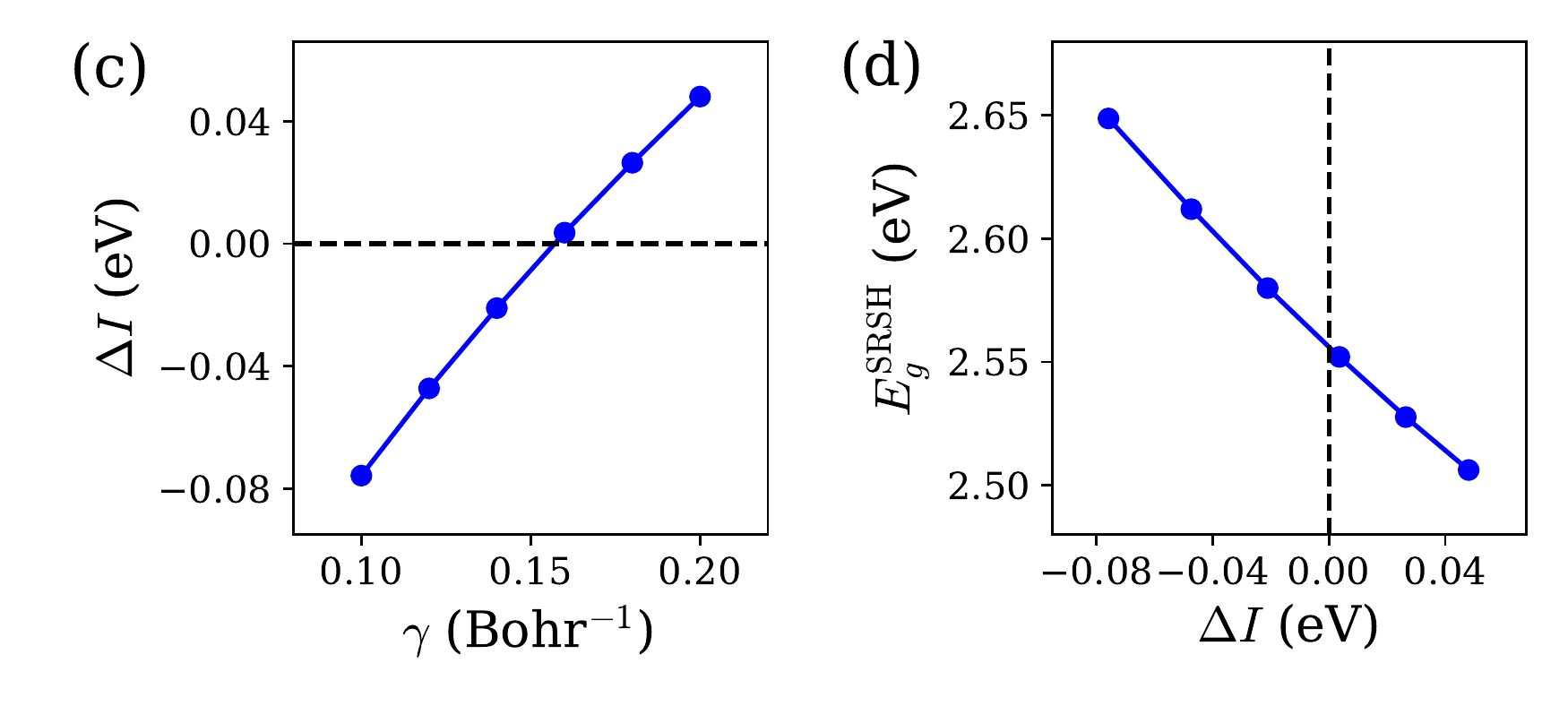}
\end{subfigure}
        \caption{\label{fig:unique_tuning_point} Illustration of the Wannier-localized, optimal-tuning SRSH approach, for the typical case of AlP. (a) Wavefunction of the valence band maximum (VBM) obtained using the PBE functional, and (b) maximally localized Wannier function used for the optimal tuning procedure. Grey - Al atoms. Purple - P atoms. Wavefunction isosurface shown in blue (positive values) and yellow (negative values) for values of $\pm 5.5\times10^{-4}$ for the VBM and $\pm 2.8 \times 10^{-4}$ for the Wannier function. (c) Deviation from the IP theorem, $\Delta I$, for the SRSH functional, as a function of the range separation parameter, $\gamma$, for AlP. (d) The fundamental band gap of AlP, calculated by SRSH, as a function of $\Delta I$. Dashed lines correspond to $\Delta I$=0.}
\end{figure*}

To demonstrate how the above approach works in practice, we use AlP as a typical example (see Fig.\ \ref{fig:unique_tuning_point}). In Fig.\ \ref{fig:unique_tuning_point}(a), we show the KS wavefunction corresponding to the valence band maximum, which is clearly delocalized over the entire supercell. In contrast, the maximally localized Wannier function (Fig.\ \ref{fig:unique_tuning_point}(b)), used in the optimal tuning procedure, corresponds to a well-defined spatial location (which specific one it is, out of all symmetry-equivalent locations, is of no consequence). In Fig.\ \ref{fig:unique_tuning_point}(c) we plot $\Delta I$ as a function of $\gamma$ for AlP. Clearly, $\Delta I$ varies monotonically with $\gamma$, such that there exists a $\gamma$ for which $\Delta I=0$. In Fig.\ \ref{fig:unique_tuning_point}(d), we plot the band gap as a function of $\Delta I$ and again find a monotonic dependence. The same behavior has been observed for all materials investigated in this work (see Table~\ref{tab:results} for the lattice parameters and WOT-SRSH functional parameters used in this study).

\section*{Results and Discussion}

\begin{SCfigure*}[\sidecaptionrelwidth][t]
\centering
\includegraphics[scale=0.88]{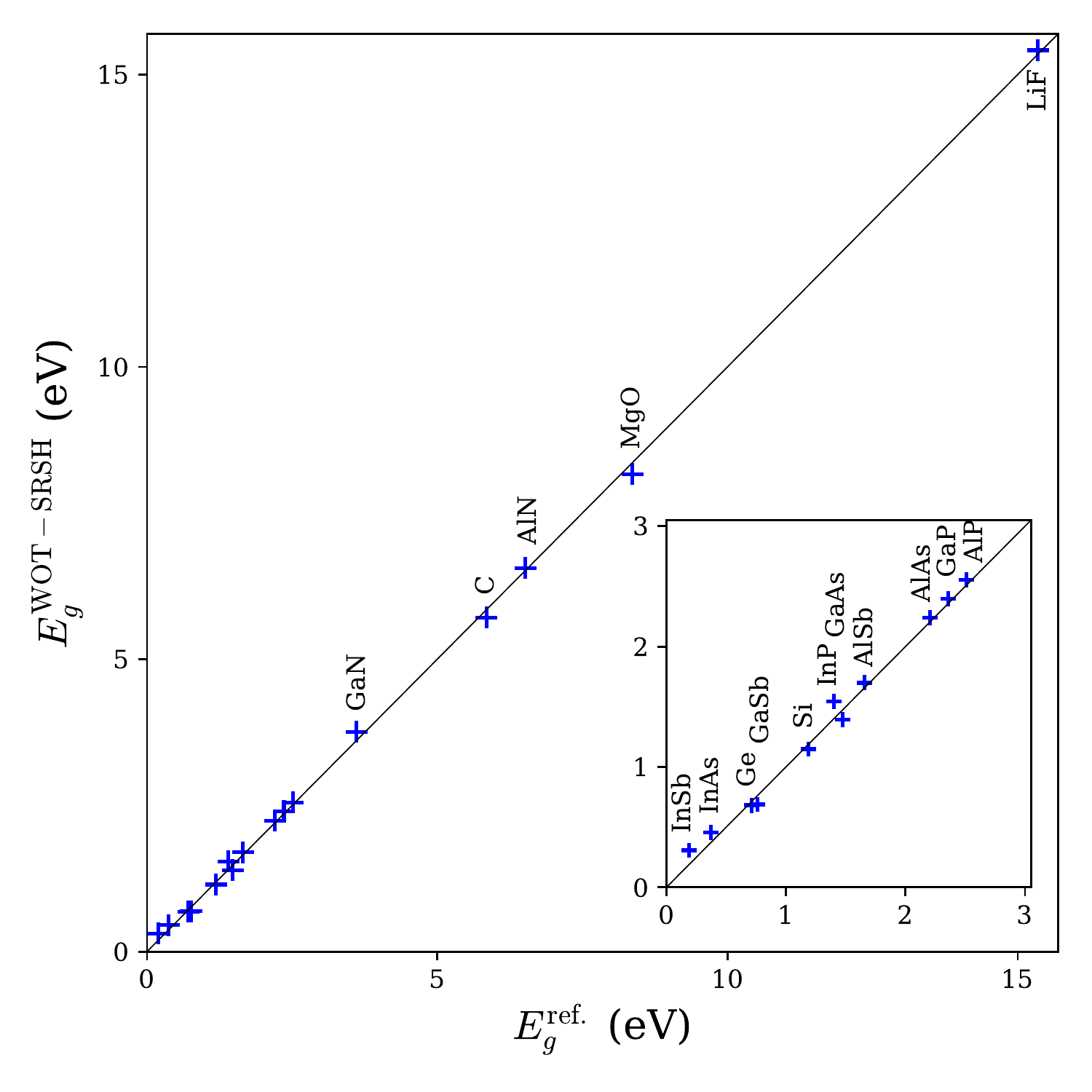}
\caption{\label{fig:band_gaps} Fundamental band gaps predicted by WOT-SRSH, compared to the reference band gaps (fundamental experimental band gaps plus zero-point renormalization energy). The straight line indicates perfect agreement. Inset: zoom-in on the 0 to 3 eV region.}
\end{SCfigure*}

A summary of the fundamental band gaps predicted using the WOT-SRSH approach, compared to experimental room-temperature fundamental band gaps\footnote{The experimental fundamental band gaps reported are produced by adding estimated or calculated exciton binding energies to the optical absorption edge or by inferring the fundamental band gap position based on the location and identification of excitonic absorption peaks.\label{fn:exp_band}}, is given in Table \ref{tab:results} and in Fig.\ \ref{fig:band_gaps}. Our calculations do not include electron-phonon coupling. Since zero point renormalization (ZPR) typically closes the band gap with respect to the fixed ion approximation \cite{giustino_cohen_2010}, we add the ZPR energy to the experimental band gaps to form a ``reference band gap'' that electronic structure theory should ideally match. Additionally, we partially account for finite temperature effects by using room temperature experimental lattice parameters. It is readily observed that excellent agreement between predicted and reference band gaps is found throughout. Importantly, the mean absolute error (MAE) is a satisfyingly small $\sim$0.1 eV, with the largest error being $\sim$0.2 eV, for MgO. Notably, for the well studied semiconductors a literature survey of the experimental band gaps reveals a mean standard deviation of $\sim$0.06 eV. This is even larger for wide band gap insulators, for example the experimental LiF band gap has an uncertainty of $\pm 0.2$ eV \cite{piacentini_1975}. Furthermore, we estimate that the overall numerical error when calculating $\Delta I$ is $\sim$0.05 eV-0.1 eV, leading to an equivalent error in the predicted band gaps (see the the slope in Fig. \ref{fig:unique_tuning_point}(b)). Thus, an MAE of $\sim$0.1 eV indicates excellent agreement for all practical purposes.

\begin{table*}[hbtp!]
\centering
\caption{\label{tab:results} Parameters of WOT-SRSH calculations and the resulting predicted fundamental band gaps compared to reference band gaps}
\begin{tabular}{c c c c c c c c}
 & $a_\textrm{lat}$ (\AA) &$\alpha$ &$\gamma$ (Bohr$^{-1}$)& $\epsilon_{\infty}$& $E_{g}^\textrm{WOT-SRSH}$ (eV) & $E_{g}^\textrm{ref.}$  (eV)& ($E_{g}^\textrm{expt.}$ , ZPR) (eV) \\
 \midrule
InSb  & 6.48\textsuperscript{a} & 0.25 & 0.17 & 13.24 & 0.3* & 0.2 & (0.17\textsuperscript{a}, 0.02\textsuperscript{b})  \\
InAs   & 6.06\textsuperscript{a} & 0.25 & 0.16 & 11.40 & 0.5* & 0.4 & (0.35\textsuperscript{a}, 0.02\textsuperscript{b})\\
Ge  & 5.66\textsuperscript{c} & 0.25 & 0.19 & 14.79 & 0.7* & 0.7 & (0.66\textsuperscript{c}, 0.05\textsuperscript{b})\\ %
GaSb & 6.10\textsuperscript{a} & 0.25 & 0.19 & 13.04 & 0.7* & 0.8 & (0.73\textsuperscript{a}, 0.03\textsuperscript{b})\\
Si & 5.43\textsuperscript{c} & 0.25 & 0.24 & 11.25 & 1.1 & 1.2 & (1.12\textsuperscript{c}, 0.06\textsuperscript{b})\\
InP  & 5.87\textsuperscript{a} & 0.25 & 0.23 & 8.87 & 1.5* & 1.4 & (1.35\textsuperscript{a}, 0.05\textsuperscript{b})\\
GaAs & 5.65\textsuperscript{a} & 0.25 & 0.15 & 10.52 & 1.4* & 1.5 & (1.42\textsuperscript{a}, 0.05\textsuperscript{b})\\
AlSb & 6.14\textsuperscript{a} & 0.25 & 0.14 & 9.82 & 1.7* & 1.7 & (1.61\textsuperscript{a}, 0.04\textsuperscript{b})\\
AlAs  & 5.66\textsuperscript{a} & 0.25 & 0.10 & 8.19 & 2.3* & 2.2 & (2.16\textsuperscript{a}, 0.04\textsuperscript{b})\\
GaP & 5.45\textsuperscript{a} & 0.25 & 0.21 & 8.89 & 2.4* & 2.4 & (2.27\textsuperscript{a}, 0.08\textsuperscript{b})\\
AlP & 5.47\textsuperscript{a} & 0.25 & 0.16 & 7.29 & 2.6 & 2.5 & (2.49\textsuperscript{a}, 0.02\textsuperscript{b})\\
GaN & 3.19, 5.19\textsuperscript{c} & 0.30 & 0.24 & 5.03 & 3.8 & 3.6 & (3.44\textsuperscript{a}, 0.17\textsuperscript{b})\\
C & 3.57\textsuperscript{c} & 0.30 & 0.23 & 5.55 & 5.7 & 5.8 & (5.47\textsuperscript{d}, 0.38\textsuperscript{e})\\
AlN & 3.11, 4.98\textsuperscript{c} & 0.35 & 0.26 & 4.12 & 6.6 & 6.5 & (6.14\textsuperscript{a}, 0.38\textsuperscript{e})\\
MgO & 4.22\textsuperscript{c} & 0.25 & 1.50 & 2.90 & 8.2 & 8.4 & (7.83\textsuperscript{f}, 0.53\textsuperscript{g})\\ 
LiF & 4.03\textsuperscript{h} & 0.25 & 1.08 & 1.93 & 15.4 & 15.3 & (14.20\textsuperscript{i}, 1.15\textsuperscript{g})\\
MAE (eV)& & & & & 0.08 &\\
MSE (eV)& & & & & 0.01 &\\
 \bottomrule 
\end{tabular}
\addtabletext{\newline Experimental room-temperature lattice parameters, $a_{\textrm{lat}}$ ($a$, $c$ for Wurtzite structure); fraction of short-range exact exchange, $\alpha$; range-separation parameter, $\gamma$; the calculated orientationally-averaged high-frequency dielectric constant, $\epsilon_\infty$; predicted band gap (* = including spin-orbit coupling effects); reference band gap (sum of the experimental room temperature fundamental band gap$\ddagger$ and the zero point renormalization energy). Also given are the mean absolute error (MAE) and the mean signed error (MSE), $E_g^{\textrm{WOT-SRSH}} - E_g^{\textrm{ref.}}$.
\newline
\textsuperscript{a}Ref.~\cite{vurgaftman2001band};
\textsuperscript{b}Ref.~\cite{cardona_thewalt_2005} Deduced by comparing the experimental 4K band gap to an extrapolated band gap;
\textsuperscript{c}Ref.~\cite{Madelung_2004};
\textsuperscript{d}Ref.~\cite{clark_harris_1964};
\textsuperscript{e}Ref.~\cite{ponce_gonze_2015} Density functional perturbation theory (DFPT);
\textsuperscript{f}Ref.~\cite{whited_walker_1973} measured at 80K; 
\textsuperscript{g}Ref.~\cite{nery_gonze_2018, chen_pasquarello_2018} GW calculation including both Migdal-Fan and Debye-Waller contributions.;
\textsuperscript{h}Ref.~\cite{landolt_bornstein_alkal_halides};
\textsuperscript{i}Ref.~\cite{piacentini_1975} measured at 200-250K.
}
\end{table*}

The results of Table~\ref{tab:results} indicate that using a fixed range-separation parameter value of $\gamma =0.2$ Bohr$^{-1}$ is a reasonable compromise for standard semiconductors. However, larger gap insulators show that this value is by no means universal. For example, for LiF we would then obtain a band gap of 13.6 eV, which is lower by a very significant 1.7 eV than the reference value reported in Table~\ref{tab:results}. Similarly, for MgO, we would obtain a gap of 7.6 eV, which is 0.8 eV lower than the reference value. These results further emphasize the importance of determining a system-specific range-separation parameter through non-empirical means, as we have outlined in this work. More details about the need for a material-dependent range-separation parameter are given in Section IV of the supplementary information \cite{SM}.

Because the WOT-SRSH method adopts the IP theorem \textit{anastz} used in the Wannier-Koopmans method (WKM) \cite{ma_wang_2016,weng_wang_2017,weng_wang_2020}, a brief discussion of the differences between the two methods is in order. In the WKM, the  \textit{ansatz} is used to derive post-processing correction terms to eigenvalues obtained using the local-density approximation (LDA). This results in an MAE of $\sim$0.15 eV \cite{ma_wang_2016}, which is comparable to the one obtained here. However, the band gaps of some materials, such as MgO and LiF are underestimated by 0.4 eV \cite{ma_wang_2016} and 1.0 eV \cite{weng_wang_2017}, respectively. Beyond the additional accuracy in band gap values, a core  strength of our approach is that because it is within the framework of GKS theory, it can be applied to any material property. For example, in the case of time-dependent DFT, because WOT-SRSH includes screened long-range exact exchange, we automatically obtain the correct long-wavelength behavior of the linear-response kernel, which has been shown to lead to optical absorption spectra on par with the Bethe-Salpeter equation (BSE) \cite{wing_kronik_2019}. Additionally, based on other studies investigating the enforcement of the IP theorem \cite{miceli_pasquarello_2018,lorke_frauenheim_2020, wing_kronik_2020}, we expect that WOT-SRSH functional parameters will be transferable to chemically similar systems, such as systems with point defects. Moreover, SRSH functionals empirically fitted to reproduce the band gap have already been shown to retain the accuracy of other hybrid functionals, including lattice parameters and vibrational frequencies \cite{seidl_egger_2021}, and we expect this to hold true for WOT-SRSH as well.

Finally, we compare the WOT-SRSH approach with many-body perturbation theory in the GW approximation, where the self-energy operator is approximated by G, the single-particle Green’s function, and W, the screened Coulomb interaction \cite{hedin_1965, hybertsen_louie_1985,hybertsen_louie_1986}. GW is perhaps the most popular approach for calculating band gaps. Importantly, one {\it could} view the screened range-separated Fock operator in WOT-SRSH as an approximate self-energy operator, which neglects the temporal dependence of the self-energy \cite{gygi_baldereschi_1989,hybertsen_louie_1986,kang_hybertsen_2010} and uses a model dielectric function \cite{gygi_baldereschi_1989,chen_pasquarello_2018}. However one does not {\it have} to view WOT-SRSH as an approximate GW scheme. Instead, as explained above the screened range-separated Fock operator is rigorously justified from generalized Kohn-Sham theory, which shows that no time-dependence is in fact needed in order to obtain accurate band gaps. We also point out that we only use a fraction $\alpha$ of exact exchange in the short range, rather than $\alpha=1$ as dictated by a model dielectric function \cite{chen_pasquarello_2018}. In order to quantitatively compare WOT-SRSH to many-body perturbation theory, we examine one GW approach, the commonly used ``single-shot'' G$_0$W$_0$ based on a PBE starting point (G$_0$W$_0$@PBE) \cite{golze_rinke_2019,hybertsen_louie_1985}. Even with such a choice, there is a spread of results in the literature due to various additional choices, e.g., plasmon pole models, pseudopotentials, basis sets, and degree of convergence \cite{van_setten_rinke_2015}. With this caveat in mind, Ref.\ \cite{wing_kronik_2019} reports that the MAE for band gaps calculated by G$_0$W$_0$@PBE, for 7 prototypical semiconductors, is 0.15 eV. The MAE of WOT-SRSH for the same set is 0.06 eV. Of note, G$_0$W$_0$@PBE underestimates the GaAs band gap by 0.5 eV, an error that is significantly larger than the error in the WOT-SRSH band gap for any material in this study. Another study \cite{van_setten_hautier_2017} reported band gaps for most of the materials in this study, with an MAE of 0.5 eV, which is significantly larger than that of WOT-SRSH. Experience with SRSH functionals which were empirically fit to match the G$_0$W$_0$@PBE band gap \cite{wing_kronik_2019} shows that the predicted position of low-lying valence bands sometimes differ from experimental values \cite{Malone_Cohen_2013}. Thus, it may be advantageous to use WOT-SRSH as a natural and well-motivated starting point for many-body perturbation theory calculations within the GW approximation, which may address this issue, as well as other properties beyond the band gap itself. 

In conclusion, we have developed a method for non-empirical selection of the parameters in a screened range-separated hybrid functional, which allows for accurate prediction of fundamental band gaps from narrow gap semiconductors to wide gap insulators entirely within density functional theory. The procedure involves optimal tuning by means of enforcing a generalized IP theorem \textit{ansatz} for localized orbitals, in this case maximally-localized Wannier functions. Practically, it only requires modest supercell calculations with a hybrid functional. It may therefore serve as a useful means not only for the prediction of band gaps, but also for the non-empirical prediction of other properties for which hybrid functionals are useful, such as optical absorption spectra and defect energetics.

\matmethods{
For convenience we calculate $\epsilon_\infty$ in step 1 with the Vienna \textit{ab initio} simulation package (\textsc{VASP}) \cite{Kresse_1996}, a plane wave code, using PBE-based projector-augmented waves (PAWs) for treating core electrons \cite{Kresse_1999}. We use an in-house modified version of the \textsc{Quantum Espresso} \cite{giannozzi_2017} plane-wave code to calculate steps 2-4. We use optimized norm-conserving (NC) Vanderbilt pseudopotentials \cite{hammann_2013} obtained from the online repository, \textsc{pseudo-dojo} \cite{van_setten_rignanese_2018} (see SI for complete computational details \cite{SM}). Methods which include exact exchange are known to be sensitive to the number of semicore states \cite{tiago_ismail-beigi_louie_2004}. We find that for Ge, Ga, In, As and Sb it is important to include one complete shell of semicore states as valence electrons. Maximally localized Wannier functions are generated using the \textsc{Wannier90} software package \cite{mostofi_marzari_2014}.

}

\showmatmethods{} 

\acknow{This work was supported via a US-Israel National Science Foundation - Binational Science Foundation (NSF-BSF) grant, DMR-1708892, and by the Israel Ministry of Defense. Computational resources were provided by the National Energy Research Scientific Computing Center, DOE Office of Science User Facilities supported by the Office of Science of the U.S. Department of Energy under Contract No. DE-AC02-05CH11231. Additional computational resources were provided by the Extreme Science and Engineering Discovery Environment (XSEDE) supercomputer Stampede2 at the Texas Advanced Computing Center (TACC) through the allocation TG-DMR190070.}

\showacknow{} 

\bibliography{pnas-sample}

\end{document}


\title{ Supplementary information to: Band gaps of crystalline solids from Wannier-localization based optimal tuning of a screened range-separated hybrid functional}

\author{Dahvyd Wing}
 \affiliation{Department of Materials and Interfaces, Weizmann Institute of Science, Rehovoth 76100, Israel}

\author{Guy Ohad}
 \affiliation{Department of Materials and Interfaces, Weizmann Institute of Science, Rehovoth 76100, Israel}

\author{Jonah B. Haber}%
\affiliation{Department of Physics, University of California, Berkeley, Berkeley, California 94720, USA }%
\affiliation{Materials Sciences Division, Lawrence Berkeley National Laboratory, Berkeley, California 94720, USA}

\author{Marina R. Filip}%
\affiliation{ Department of Physics, University of Oxford, Oxford OX1 3PJ, United Kingdom }%

\author{Stephen E. Gant}%
\affiliation{Department of Physics, University of California, Berkeley, Berkeley, California 94720, USA }%
\affiliation{Materials Sciences Division, Lawrence Berkeley National Laboratory, Berkeley, California 94720, USA}

\author{Jeffrey B. Neaton}
\affiliation{Department of Physics, University of California, Berkeley, Berkeley, California 94720, USA }
\affiliation{Materials Sciences Division, Lawrence Berkeley National Laboratory, Berkeley, California 94720, USA}
\affiliation{Kavli Energy NanoSciences Institute at Berkeley, Berkeley, CA, 94720, USA }

\author{Leeor Kronik}
 \affiliation{Department of Materials and Interfaces, Weizmann Institute of Science, Rehovoth 76100, Israel}

\date{\today}

\maketitle

\section{Computational details}

In table \ref{tab:cutoff_grid} we show the numerical parameters used for each step of obtaining the WOT-SRSH functional, as described in the main text. For step 1 we calculate the ion-clamped, high-frequency dielectric constant, $\epsilon_\infty$, by the change in polarization in response to a finite electric field \cite{Nunes_Gonze_2001,Souza_Vanderbilt_2002}. In these calculations local field effects are included for both Hartree and exchange-correlation potentials \cite{northrup_louie_1987}. The Ga, Ge, and In PAWs include semicore $d$-electrons as valence ones. The total energy convergence criteria for calculations where the small electric field is applied is $10^{-10}$ eV. Most of the dielectric constants used for this study were already reported in Refs.~\cite{wing_kronik_2019,wing_kronik_2020_insb}.

For steps 2-4, we use version 0.4 pseudopotentials from \textsc{pseudo-dojo} \cite{van_setten_rignanese_2018}. For most pseudopotentials we use stringent fully relativistic pseudopotentials. We use standard scalar relativistic pseudopotentials for Si, AlP, AlN, C, MgO, and LiF. No projections were specified to generate a starting point for Wannierization, rather Bloch wavefunctions, without unitary transformation, were used for the starting point.

\begin{table*}[hbp!]

\centering
\caption{\label{tab:cutoff_grid} Plane wave energy cutoff ,$E_\textrm{cutoff}$, and $k$-grid size for calculating the dielectric constant, $\epsilon_\infty$; deviation from satisfaction of the generalized ionization potential theorem, $\Delta I$; and the fundamental band gap, $E_g^{\textrm{WOT-SRSH}}$}
\begin{tabular}{ c c c c c c}
\hline
\hline
&  \multicolumn{2}{c}{$\epsilon_\infty$} & $\Delta I$ & \multicolumn{2}{c}{$E_g$} \\
\hline
 &$E_\textrm{cutoff}$ (eV) & $k$-grid & $E_\textrm{cutoff}$ (Ry)& $E_\textrm{cutoff}$ (Ry) & $k$-grid \\
InSb  &300 &$8\times8\times8$  & 70 & 80 & $8\times8\times8$ \\
InAs& 300  & $8\times8\times8$  & 100 & 110 & $8\times8\times8$ \\
Ge&  300& $9\times9\times9$  &110 & 120 & $10\times10\times10$\\ 
GaSb& 300&$8\times8\times8$  & 100 & 110 & $8\times8\times8$\\
Si& 300 & $11\times 11\times 11$ & 20 & 30 & $10\times10\times10$\\
InP& 300 & $11\times 11\times 11$ & 60 & 70 & $8\times8\times8$\\
GaAs&300 & $11\times 11\times 11$ & 100 & 110 & $8\times8\times8$\\
AlSb&300 & $11\times 11\times 11$ & 75 & 85 & $8\times8\times8$\\
AlAs& 300&$11\times 11\times 11$ & 95 & 105 & $8\times8\times8$\\
GaP& 300 &$11\times 11\times 11$ & 90 & 100 & $8\times8\times8$\\
AlP& 300  &$11\times 11\times 11$ & 30 & 40 & $8\times8\times8$\\
GaN&450 &$8\times8\times5$ & 95 & 105 & $9\times9\times6$\\
C &450 &$9\times 9\times 9$ & 70 & 80 & $8\times8\times8$\\
AlN&450 &$6\times6\times4$ & 60 & 70 & $9\times9\times6$\\
MgO&600  &$4\times 4 \times 4$ & 70 & 100 & $8\times8\times8$\\ 
LiF& 450 &$4\times4\times4$ & 100 & 110 & $8\times8\times8$\\
\hline
\hline
\end{tabular}

\end{table*}

\section{Convergence of $\Delta I$ with respect to supercell size}
\begin{figure}[htbp!]
\includegraphics[scale=0.75]{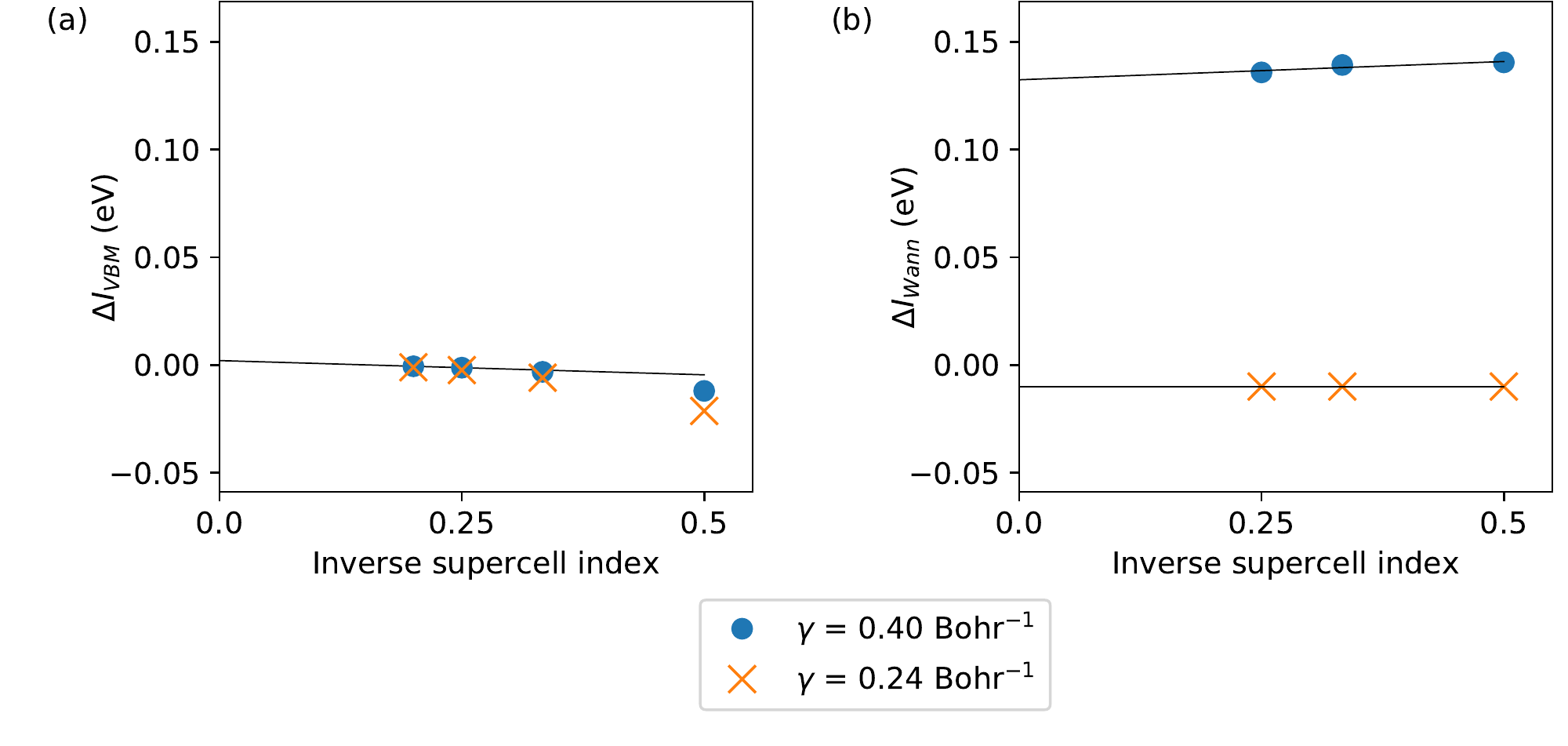}
\caption{\label{fig:convergence} $\Delta I$ as a function of the inverse supercell index ($1/n$ for an $n{\times} n{\times} n$ supercell), using SRSH functionals with $\alpha=0.25$, $\epsilon=11.25$, and two different values of $\gamma$, for electron removal from (a) the VBM, (black line: a linear extrapolation based on the two largest supercell sizes), and (b) a Wannier function using Eq.\ (4) of the main text (black lines: linear fit).}
\end{figure}

Fig.\ \ref{fig:convergence} shows $\Delta I$ for Si as a function of supercell size, using SRSH functionals with two different values of $\gamma$. Fig.\ \ref{fig:convergence}(a) shows $\Delta I$ when removing an electron from the VBM maximum, without any localization scheme, i.e. $\Delta I_{VBM} = E[N-1] - E[N] +\epsilon_{VBM} + \frac{\alpha_{mad}q^2}{2 \epsilon_\infty L} $, where the last term is the image charge correction discussed in the main text, which facilitates convergence \cite{broqvist_pasquarello_2009}. For computational convenience, this is calculated using VASP with a planewave cutoff of 250 eV. Clearly, $\Delta I_{VBM}$ approaches 0 in the limit of large supercells, regardless of the value of $\gamma$. Fig.\ \ref{fig:convergence}(b) shows $\Delta I$ for the same two SRSH functionals, but now upon removing an electron from a Wannier function generated using PBE, as per Eq.\ (4) of the main text. Clearly, $\Delta I$ is nearly converged with respect to supercell size even for a modest $2\times 2\times 2$ supercell. Furthermore, SRSH functionals with different values of $\gamma$ converge to different values of $\Delta I$, which provides a basis for selection of the optimal $\gamma$. 

Next, we examine the convergence of $\Delta I$ for seven materials, for different supercell sizes, while keeping the SRSH parameters fixed to the values summarized in Table \ref{tab:supercell}. The results for Si, AlP, InSb, MgO and LiF show that $\Delta I$ for a $2\times2\times2$ supercell is sufficiently converged. For AlN and GaN, it was prohibitively expensive to calculate supercells larger than $3\times3\times2$. Based on the convergence of the other materials we estimate that $\Delta I$ is converged to 0.1 eV for AlN and GaN. 

\begin{table*}[tpbh!]
\centering
\caption{\label{tab:supercell} $\Delta I$ (eV) as a function of supercell size for selected materials}
\begin{tabular}{c c c c c c}
\hline
\hline
Supercell size      & Si     & AlP    & InSb   & MgO    & LiF   \\
${1\times1\times1}$ & -0.20 & -0.13  & -0.33 & 0.09  & -0.05  \\
${2\times2\times2}$ & -0.01  & -0.02 & 0.00  & 0.02 & 0.01\\
${3\times3\times3}$ & -0.01 & -0.02 & 0.02  & 0.02  & 0.01  \\ 
${4\times4\times4}$ & -0.01 & -0.02 &        &        &         \\
\hline
& AlN & GaN \\
${2\times2\times1}$ & 0.07 & -0.03 & \\
${3\times3\times2}$ & 0.09 & 0.04 & \\
\hline
\hline
\end{tabular}

\end{table*}

The Makov-Payne image charge correction has been shown to accelerate convergence of $\Delta I$ calculations with respect to the supercell size \cite{wing_kronik_2020}. We use the same value of the dielectric constant in the Makov-Payne correction in Eq. (4) as we use for setting the fraction of the screened long-range exact exchange in the SRSH functional. This dielectric constant is converged with respect to the $k$-grid and includes spin-orbit coupling effects in order to be consistent with the calculation of the band gap of the material in step 4. We find that using Makov-Payne corrections with dielectric constants calculated without spin-orbit effects and with a $k$-grid that matches the supercell used in the $\tilde{E}_{\textrm{constr}}^\gamma[\phi](N-1)$ calculations causes $\Delta I$ calculations to converge somewhat slowly with respect to the supercell size, while ultimately yielding the same result. 

\section{Selecting $\alpha$}
\begin{figure}[htbp!]
\includegraphics[scale=0.64]{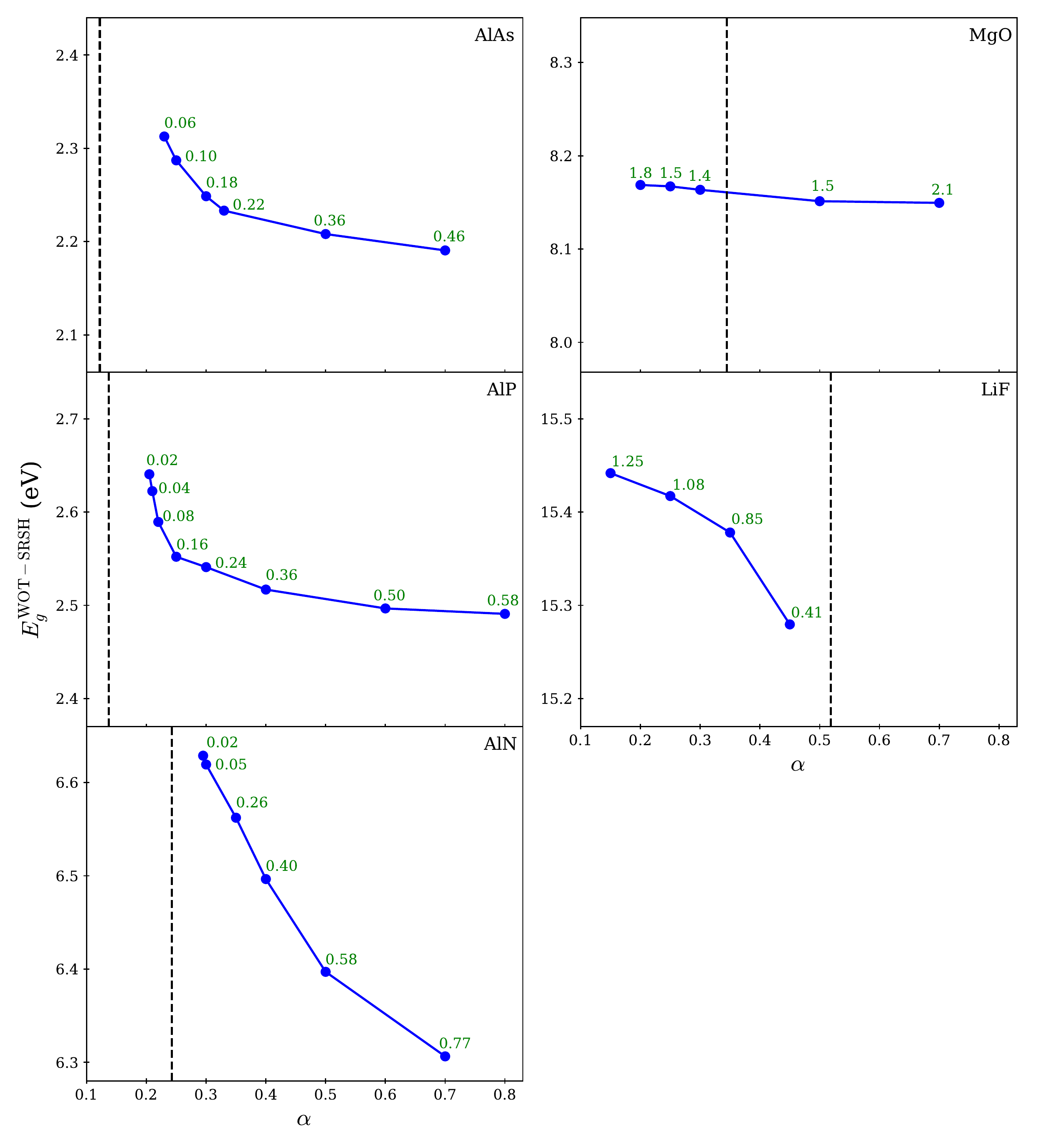}
\caption{\label{fig:alpha} Band gap predicted by the WOT-SRSH functional, as a function of $\alpha$. The tuned value of $\gamma$ is reported above each point in Bohr$^{-1}$. The vertical dashed line is the value of $\frac{1}{\epsilon_\infty}$ and indicates a region of values of $\alpha$ where, in general, one cannot find a reliable optimal value of $\gamma$.}
\end{figure}

While we use a default $\alpha$ value of 0.25, this value may prohibit optimal tuning in some cases. To understand why, consider the case of  $\alpha = \frac{1}{\epsilon}$. Then $\Delta I$ would be independent of $\gamma$, even upon using the IP theorem \textit{ansatz}. Even if $\alpha$ is not exactly equal to $\frac{1}{\epsilon}$, poor numerical sensitivity, or failure to satisfy the IP theorem {it ansatz} at all, may ensue. In such cases, we increase $\alpha$ until we can obtain a unique $\gamma$ that does not approach zero. That is because the case of a very small $\gamma$, known as the “$\gamma$ collapse problem” \cite{de_queiroz_kuemmel_2014,Bhandari__dunietz_2018}, corresponds to an exceedingly large screening length so the functional is effectively “all short range”, which is not physical. Therefore $\frac{1}{\gamma}$ should not exceed the supercell size. To show explicitly that this procedure is legitimate, and does not introduce undue ambiguity, we investigated the prediction of WOT-SRSH using different values of $\alpha$ for AlAs, AlP, AlN, MgO, and LiF (see Fig.\ \ref{fig:alpha}). We found that for all materials, except AlN, choosing an $\alpha$ in the range of 0.15 to 0.5 changes the band gap by a minimal 0.1 eV. For AlN, one can satisfy $\Delta I=0$ only for $\alpha>0.3$ and then $\alpha$ in the range of 0.3 to 0.5 changes the band gap by only 0.2 eV. 

\section{Why a material-independent value of $\gamma$ is not sufficiently accurate}

\begin{figure}[htbp!]
\includegraphics[scale=0.7]{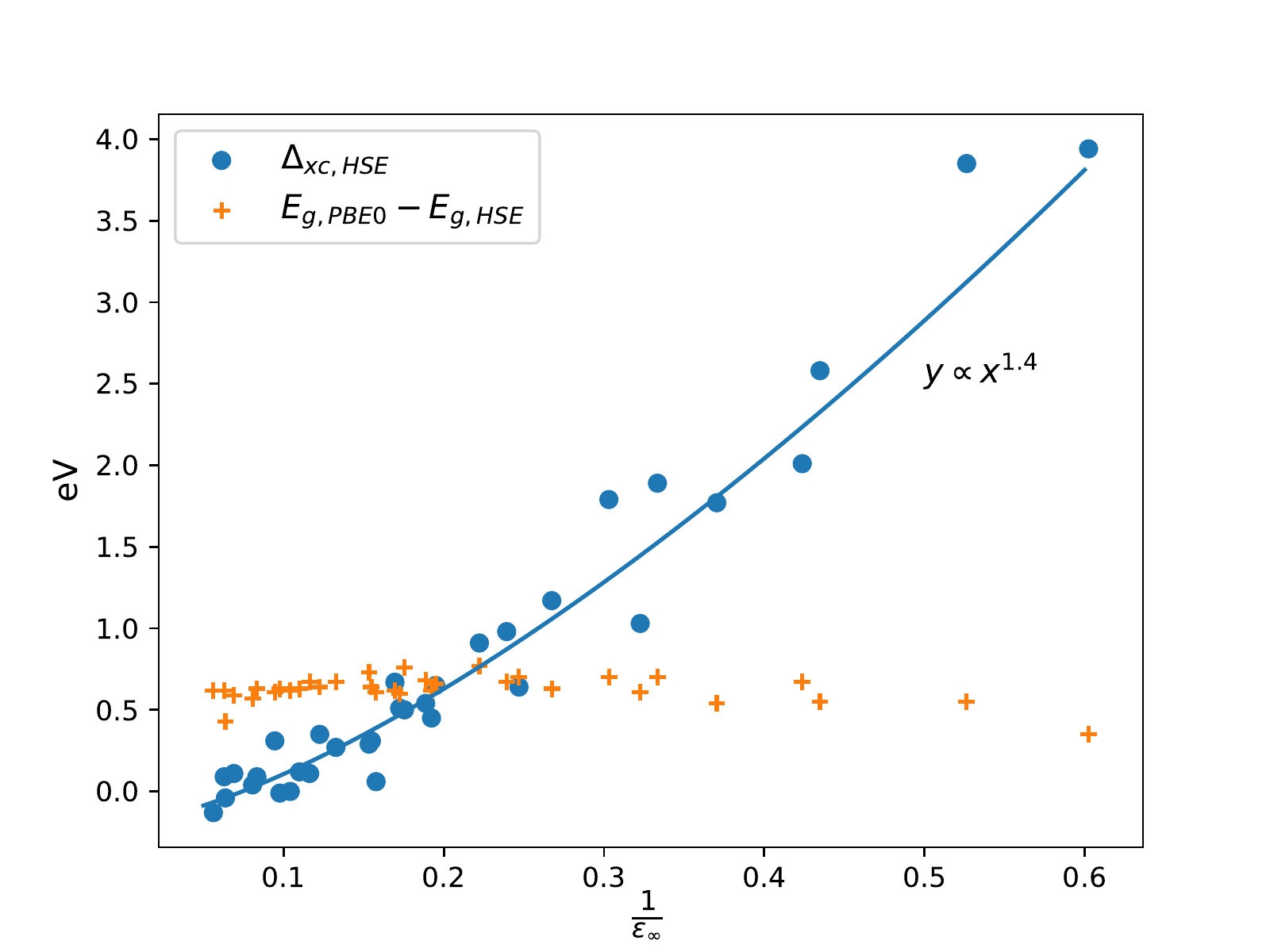}
\caption{\label{fig:nonlinear_delta_xc}  An estimation of the missing derivative discontinuity of HSE, $\Delta_{xc,HSE}$ (based on the difference between the experimental gap and the HSE gap), with respect to the experimental high-frequency dielectric constant of the material (blue circles). Data for GaSb, InAs, and InSb are taken from Ref.~\cite{wing_kronik_2020_insb} and data for all other materials are taken from Ref.~\cite{chen_pasquarello_2018}. The amount that 25\% long-range exact exchange increases the band gap for the same set of materials is given in orange plus signs. Here, the HSE and PBE0 data were taken from Ref.~\cite{borlido_botti_2019}. }
\end{figure}

As mentioned in the main text, a fixed value of $\gamma =0.2$ Bohr$^{-1}$ yields accurate band gaps for the narrow to medium gap semiconductors investigated in this study, but not for wide gap insulators. Here we present three observations that, taken together, explain why a fixed value of $\gamma$ will not produce accurate band gaps for all materials. The first observation is that for a hybrid functional with a fixed parameterization of the Fock operator, the missing derivative discontinuity $\Delta_{xc}$ depends superlinearly on $1/\epsilon_\infty$, where $\epsilon_\infty$ is the high-frequency dielectric constant. This can be seen for the HSE functional \cite{Heyd_Ernzerhof_2003,*heyd_ernzerhof_2006} in Fig. \ref{fig:nonlinear_delta_xc}, where $\Delta_{xc,HSE}$ is estimated by $\Delta_{xc,HSE}\approx E_{g,ref}-E_{g,HSE}$, i.e. the difference between the experimental gap (corrected for ZPR) and the HSE gap. We find that the derivative discontinuity increases as $\propto(1/\epsilon_\infty)^{1.4}$. The second observation is that a material-independent parameterization of long-range exact exchange increases the band gap by approximately the same amount for different materials. This can be seen in the scatter plot of the difference between the PBE0 band gap and the HSE band gap in Fig. \ref{fig:nonlinear_delta_xc}. Quantitatively, we find that 25\% long-range exact exchange increases the gap by $\sim$0.6 eV. The third observation is that, for a fixed value of $\gamma$, the band gap increases linearly with the fraction of long-range exact exchange, see for example Ref.\ \cite{zheng_cororpceanu_2017}. Together, this means that for any fixed value of $\gamma$, a fraction $\frac{1}{\epsilon_\infty}$ of long-range exact exchange will increase the band gap linearly for different materials, with respect to the materials dielectric constant, whereas the missing derivative discontinuity, $\Delta_{xc}$, increases superlinearly. Therefore, for a fixed value of $\gamma$, calculated band gaps can be relatively accurate for {\it some} range of dielectric constants, but not for the {\it entire} range. Even if we ignore this issue, Fig.\ \ref{fig:nonlinear_delta_xc} shows that there is enough variation from material to material, independent of the dielectric constant, such that it is still preferable to tune $\gamma$ per material, rather than to construct, say, an empirically fitted function, $\gamma(\epsilon_\infty)$.

\bibliography{pnas-sample.bib}
\bibliographystyle{apsrev4-2_new.bst}